\documentclass[11pt]{IEEEtran}
\IEEEoverridecommandlockouts
\usepackage[figuresright]{rotating}
\usepackage{graphicx}
\usepackage{subfigure}
\usepackage{cite}
\usepackage{caption}
\usepackage{multirow}
\usepackage{tabularx}
\usepackage{fancyhdr} 
\usepackage{makecell}
\usepackage{lscape} 
\usepackage{float}
\begin{document}
\title{Defense Strategies Toward Model Poisoning Attacks in Federated Learning: A Survey
}

\author{Zhilin Wang, Qiao Kang, Xinyi Zhang, Qin Hu$^*$

\thanks{This work is partly supported by the US NSF under grant CNS-2105004.}
\thanks{Zhilin Wang, Qiao Kang, and Qin Hu are with the Department of Computer and Information Science, Indiana University-Purdue University Indianapolis, Indiana, USA. E-mail:\{wangzhil, kangjoe, qinhu\}@iu.edu.}
\thanks{Xinyi Zhang is with the Department of Computer Science, Purdue University-West Lafayette, Indiana, USA. Email: zhan3652@purdue.edu.}
\thanks{$^*$ Corresponding author.}}

\maketitle

\begin{abstract}
Advances in distributed machine learning can empower future communications and networking. The emergence of federated learning (FL) has provided an efficient framework for distributed machine learning, which, however, still faces many security challenges. Among them, model poisoning attacks have a significant impact on the security and performance of FL. Given that there have been many studies focusing on defending against model poisoning attacks, it is necessary to survey the existing work and provide insights to inspire future research. In this paper, we first classify defense mechanisms for model poisoning attacks into two categories: evaluation methods for local model updates and aggregation methods for the global model. Then, we analyze some of the existing defense strategies in detail. We also discuss some potential challenges and future research directions. To the best of our knowledge, we are the first to survey defense methods for model poisoning attacks in FL.
\end{abstract}

\section{Introduction}
The rapid development of artificial intelligence (AI) has greatly changed society. Given that a large amount of data can be generated in our daily life, how to effectively and efficiently use the data to train machine learning models has become a challenge that needs to be addressed. Traditional machine learning frameworks require servers to collect data and perform training tasks in a centralized way, which causes many problems and hinders the development of AI: 1) it's expensive to collect enough data; 2) performing machine learning on servers consumes a lot of resources, such as computation, communication, and storage resources; 3) transferring original data from end devices to servers can lead to data privacy leakage.

The emergence of federated learning (FL) has provided a promising solution to tackle the above problems. Google proposed the concept of FL in 2016, which is a distributed machine learning framework\cite{mcmahan2017communication,kairouz2019advances}. The basic idea of FL is that multiple end devices, i.e., clients, collaboratively train a machine learning model. Unlike traditional machine learning frameworks, FL does not require clients to transmit raw data to a central server, but only the updates of the trained local models, thus protecting the data privacy of clients. FL is suitable for large-scale machine learning tasks because it distributes the training task to a large number of end devices, while the central server is only responsible for model aggregation, thus reducing the computational pressure on the server. Currently, FL has been applied in various fields, such as healthcare\cite{xu2020towards}, transportation\cite{zhang2021towards}, communications \cite{liu2020secure}, and Internet of the Things (IoT)\cite{lu2019blockchain}. In particular, FL has been used to support the development of 5G and 6G, which can enable more secure and efficient schemes for future communications and networking.

However, FL has also encountered several challenges. Among them, security is one of the most important concerns for researchers. Although FL does not require clients to upload raw data to protect data privacy, due to the distributed nature of FL, there is no guarantee that all devices involved in training are honest, which means that they may upload malicious submissions. In addition, end devices can be vulnerable to external attacks, leading to erroneous local model updates. There are many attacks on FL, such as poisoning attacks\cite{tolpegin2020data,cao2019understanding}, backdoor attacks\cite{bagdasaryan2020backdoor,sun2019can}, and inference attacks\cite{nasr2019comprehensive,luo2021feature}. Poisoning attacks are divided into data poisoning attacks and model poisoning attacks, which are both untargeted attacks. In other words, the aim is to make the model performance degraded generally instead of achieving some targeted misclassification. Data poisoning attacks manipulate the raw data on the clients\cite{sun2021data}, while model poisoning attacks manipulate local model updates\cite{zhou2021deep}. Some studies have shown that model poisoning attacks are more likely to cause damages to FL than data poisoning attacks\cite{cao2020fltrust}.

Though lots of existing surveys focus on analyzing the security problems faced by FL, little attention has been paid to defense methods. For example, the work in \cite{mothukuri2021survey,jere2020taxonomy,lyu2020threats} details the possible security problems of FL, but there is less analysis on how to defend against model poisoning attacks. Considering about the severity of model poisoning attacks to FL, it is necessary to survey its defense mechanisms so as to attract more attention and inspire future research.

In our survey, we first introduce the relevant background knowledge about FL and model poisoning attacks, then we classify and detail the existing defense methods, and finally, we discuss the challenges and future research directions. To the best of our knowledge, this is the first survey on the defense mechanisms of model poisoning attacks. Our main contributions are as follows.
\begin{itemize}
    \item We investigate the existing defense methods for model poisoning attacks in FL and classify them into two categories: evaluation methods for local model updates and aggregation methods of the global  model. 
    \item We describe some of the defense methods in detail, analyzing their workflows and application scenarios.
    \item We summarize the challenges of defense methods against model poisoning attacks and discuss future research directions.
\end{itemize}

The remainder of this paper is organized as follows. We introduce the background knowledge of FL and model poisoning attacks in Section \ref{back}. The detailed analysis of defense strategies toward model poisoning attacks in FL is described in Section \ref{defense}. In Section \ref{challenge}, we discuss the challenges and some promising future research directions. In the end, we conclude the whole paper in Section \ref{concl}.


\section{Background Knowledge}\label{back}

In this section, we illustrate the background of FL and model poisoning attacks.

\subsection{Federated Learning}\label{fl}

We consider a conventional FL framework,  which consists of a \textit{central server} and numerous local devices termed \textit{clients}. 
We let $i\in \left\{1,2,3,\cdots,N\right\}$ represent an individual client, where $N$ is the total number of clients. Each client has a different size of local data set, which can be denoted as $D_i$. At the beginning of each round, the server first selects a certain number of clients to participate in the federated learning, and we use $m_k$ to represent the fraction of clients chosen in round $k$, where $k\in\left\{1,2,3,\cdots,K\right\}$ and $K$ is the total number of training rounds for a specific FL task. Once the clients are selected, the server sends the initial global model, denoted as $w_0$, to those clients.
Then clients start to train local models using their own raw data based on $w_0$, and send the trained local model updates $w_i^k$ to the central server, where $w_i^k$ is the updates submitted by client $i$ in round $k$. Next, the central server collects the local model updates and runs an aggregation algorithm to update the global model. This process will be terminated once the loss of the global model is converged. The most popular aggregation algorithm is \textit{federated averaging} (FedAvg)\cite{wang2020optimizing,mcmahan2017communication}, and it can be expressed as
    $\delta^k=\frac{\sum_{i=1}^N D_{i} \delta_i^k}{\sum_{i=1}^N D_{i}},$
where $\delta^k$ is the final weight differences of all clients and $\delta_i^k$ is the weight difference of each client in round $k$, and $\delta_i^k$ can be calculated by
    $\delta_i^k = w_i^k-w_i^{k-1}$.

\subsection{Model Poisoning Attacks}
FL is a distributed learning framework that requires multiple devices to participate, but there is no guarantee that the selected devices will work honestly. In other words, the clients in FL are not trustworthy for the central server. This can lead to many potential security problems, such as poisoning attacks, backdoor attacks, and inference attacks. In this paper, we focus on the model poisoning attack, which is one of the most popular attacks against FL.

\begin{figure}[htbp]
\centering
\includegraphics[width=0.30\textwidth]{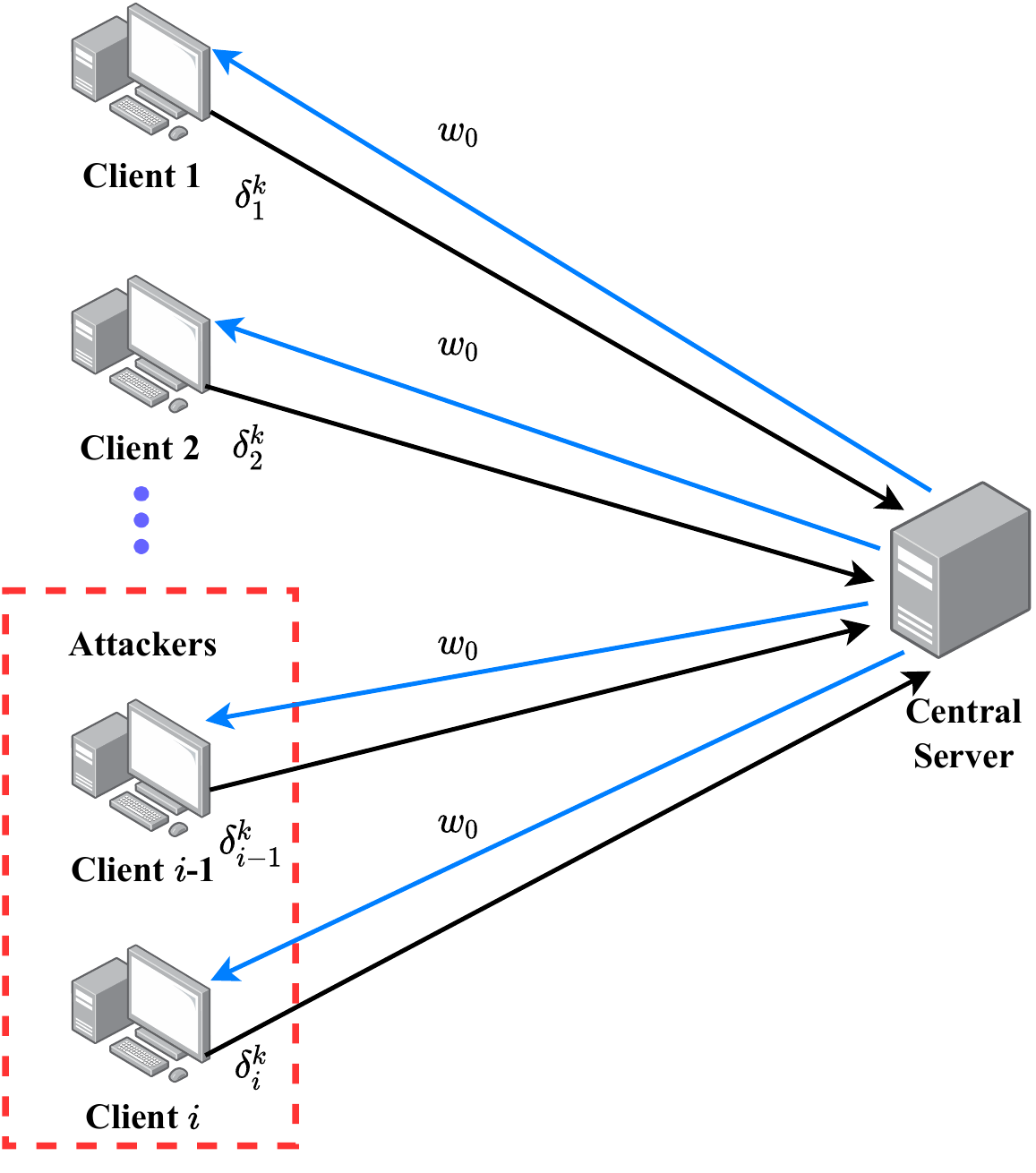}
\caption{The topology of model poisoning attacks in FL.
}
\label{fig:framework}
\end{figure}

Model poisoning attacks can be initiated by malicious clients or by an external attacker who controls some clients. In this paper, we do not distinguish the attack initiators and uniformly refer to them as malicious clients.
The topology of model poisoning attacks in FL is shown in Fig. \ref{fig:framework}.
Specifically, 
when the malicious clients finish training based on the initial global model $w_0$ in round $k$, they modify the local model updates $\delta_i^k$ and then submit them to the central server. Since the central server does not have access to the raw data of the local clients, and the data on the clients are usually non-independent and identically distributed (non-IID), it is difficult for the central server to identify the modified local updates, which leads to slow convergence or reduced accuracy of the global model. Generally speaking, model poisoning is an untargeted attack, where the purpose is not to target a specific label but the overall performance of the final model. 

Based on the number of malicious clients, we can classify the model poisoning attack as single malicious clients-initiated attacks and multiple malicious clients-initiated attacks. As for the former, its effect is more obvious when the number of clients is small; while when the number of participants is large, its impact will be offset by weights-based algorithms such as FedAvg. The latter one is more practical in the mobile scenario.

The reasons why it is difficult to detect and defend against model poisoning attacks are as below:

\begin{itemize}
    \item The data of local clients are non-IID, which leads to significant differences among the obtained local model updates, causing difficulties for detection.
    \item The central server cannot obtain the raw data of local clients, and therefore cannot use these data for verification.
    \item The current popular model aggregation method (i.e., FedAvg) relies on the data volume of clients to assign certain weights to the updates submitted by clients, which does not have any special treatment for the contaminated updates, and thus it cannot defend against model poisoning attacks.
\end{itemize}

\section{Defense Strategies Towards Model Poisoning Attacks}\label{defense}
In this section, we analyze the existing defense strategies toward model poisoning attacks. Although attack detection and defense are two different phases, we treat them as the same in this paper since they usually work together to protect FL. From the existing research, approaches to defend against model poisoning attacks can be divided into two categories: one is to identify malicious submissions by designing evaluation mechanisms for local model updates, and the other one is to design novel and Byzantine fault-tolerant aggregation algorithms based on mathematical statistics. These two approaches are usually used jointly.


\subsection{Evaluation Methods for Local Model Updates}
Intuitively, the most straightforward way to defend against model poisoning attacks is to examine the submissions of clients. However, since the central server does not have the direct access to the raw data of the end devices, evaluating the model updates submitted by devices has become a challenge. In this part, we will discuss some of the existing evaluation methods in detail.

\subsubsection{Spectral Anomaly Detection Based Method}

The basic idea of spectral anomaly detection is to embed benign data and malicious data into a low-dimensional space.
In \cite{li2020learning}, a spectral anomaly detection based evaluation framework for FL is proposed. In this framework, they first assume that there will be a public dataset which can be trained to provide the spectral anomaly detection model. Then, they embed the local model updates, including benign updates and malicious updates, into a low-dimensional latent space. In this way, the essential features of these updates are well maintained, and the two kinds of updates can be easily distinguished after removing the noisy and redundant features. After the detection process, the malicious updates are removed and only the benign updates are taken into account during the global model aggregation process. According to the experimental results, the spectral anomaly detection based evaluation method can perform well in eliminating the abnormal updates and maintaining high accuracy of the model at the same time.

\subsubsection{Truth Inference Based Evaluation Method} In \cite{tahmasebian2021robustfed}, the authors utilize an optimization based truth inference method to evaluate the reliability of the submitted updates in FL before the aggregation of the global model. The basic idea of truth inference in FL is minimizing the weighted deviation from the true aggregated parameters. First, they calculate the reliability score of each update. Then, they propose two methods to aggregate the global model: the first one is using the reliability score as the weight of each update, and the other one is to remove updates with low reliability scores. However, this paper only focuses on IID data, making it not practical for non-IID cases.

\subsubsection{Entropy Based Filtering Method} Park et al.\cite{park2021sageflow} design an entropy-based filtering scheme to detect the outlier updates. At the beginning, the server collects some public data, and then calculates the entropy of each update with the public data. Based on their experimental observations, they argue that the updates with higher entropy will lead to lower accuracy during the testing stage. Thus, they set a threshold for the entropy and filter out updates with entropy higher than the threshold. They further illustrate that the entropy-based filtering method can perform well even when the number of adversaries is large, overcoming the limitation of attack ratio.

\subsubsection{Cosine Similarity Based Evaluation Method}

Cosine similarity is defined by calculating the cosine of the angle between two vectors to evaluate their similarity.

Cao et al. \cite{cao2020fltrust} utilize cosine similarity to assess the similarity between each update and the update obtained by training based on the clean dataset of the server. They argue that an attacker can manipulate the directions of updates to achieve the purpose of model poisoning attack, and the directions of the updates can, to a certain extent, indicate the honesty of the end devices. They first let the server collect a small sample size of data (e.g., 100 samples) as the clean dataset, based on which the server trains the model. After the calculation of cosine similarity, there will be a trust score for each update used as the weight for the global model aggregation. 

In \cite{sattler2020byzantine}, the impact of the model poisoning attack is mitigated according to dividing the updates into different groups by the cosine similarity between updates submitted by clients. This is a new framework called Clustered Federated Learning. In \cite{xu2020towards}, a cosine similarity based evaluation method is applied to detect malicious updates. The central server keeps the reputation of each participant by checking the similarity of local model updates and removes non-contributing or malicious participants. Different from \cite{cao2020fltrust}, the schemes in \cite{sattler2020byzantine} and \cite{xu2020towards} require no collected clean dataset, and the cosine similarity is calculated between two different local model updates.

\subsubsection{Learned Lessons} Malicious nodes can be effectively identified by evaluating local model updates before model aggregation, thus reducing the negative impact of model poisoning attacks on FL. The evaluation methods mentioned above require examination of the data submitted by each client, which consumes a long time and computational resources. In addition, some evaluation methods require the server to collect a portion of clean data to be used as a basis for validating model updates to perform machine learning accordingly, which may lead to new problems, such as energy consumption and privacy leakage.

\subsection{Aggregation Methods for the Global Model} 
The aggregation of the global model is an important part of FL. Currently, conventional FL uses FedAvg as the aggregation method, which is unable to identify malicious submissions and leads to the success of model poisoning attacks. A number of studies have focused on designing novel aggregation algorithms to improve the robustness of FL. Based on the existing research, the aggregation methods used to defend against model poisoning attacks can be broadly classified into two categories: adjusting the weights of 
local model updates based on certain criteria and designing aggregation algorithms using statistical methods.

\subsubsection{Criteria-based Aggregation Methods}
 The metrics here refer to some criteria used to evaluate local model updates (e.g., trust, reliability, similarity), and they are derived from the examination of the updates. For example, in \cite{cao2020fltrust}, the authors use trust as the weights for local model updates in the aggregation process, while in \cite{tahmasebian2021robustfed}, the authors use the reliability of local model updates as the weights. It should be noted that some aggregation methods directly discard data that do not satisfy the criteria, which is also a way of weighting, i.e., treating the weights as 0. 

\subsubsection{Statistic-based Aggregation Methods}Different from criteria-based aggregation methods mentioned above, the statistic-based aggregation method does not perform verification of local model updates, but only selects data by statistical methods during the global model aggregation process.

\textit{Trim-mean}
\cite{yin2018byzantine} is to select each parameter of the model independently, sort and remove the maximum and minimum values, and calculate the mean value as the aggregated value of the parameter. Specifically, for each $j$-th model parameter, the server ranks the $j$-th parameter of $m$ local models, i.e., $w_{1j}$, $w_{2j}$, ..., $w
_{mj}$, where $w_{ij}$ is the $j$-th parameter of the $i$-th local model, removes the largest and smallest $\beta$ of them and calculates the average of the remaining $m-2\beta$ parameters as the $j$-th parameter of the global model. Assuming that at most $c$ clients are corrupted. This pruned average aggregation rule achieves a sequentially optimal error rate of $\widetilde{O}(\frac{c}{m\sqrt{n}}+\frac{1}{\sqrt{mn}})$ when $c \leq \beta < \frac{m}{2}$ and the objective function to be minimized is strongly convex, where $n$ is the number of training data points on the clients (with the assumption that each client has the same number of training data samples).

\textit{Median}
\cite{yin2018byzantine} is another aggregation method which selects the median value independently among the parameters as the aggregated global model. In this Median aggregation rule, for each $j$-th model parameter, the server ranks the $j$-th parameter of $m$ local models and takes the median as the $j$-th parameter of the global model. Like the Trim-mean aggregation rule, the Median aggregation rule achieves the sequentially optimal error rate when the objective function is strongly convex.

\textit{Krum}
\cite{blanchard2017machine} selects a local model among $m$ local models, which is the closest to the others, as the global model. The advantage is that even if the selected model comes from a malicious attacker, its impact may be limited because it is similar to other local models that may come from benign clients. Assume that at most $c$ clients are compromised. For each local model $w_i$, the server computes the sum of the distances between $m-c-2$ local models with the closest Euclidean distance to $w_i$. Krum selects the local model with the smallest sum of distances as the global model. When $c<\frac{m-2}{2}$, Krum has theoretical guarantees for convergence of certain objective functions.

\textit{Bulyan}
\cite{guerraoui2018hidden} is a Byzantine fault-tolerant algorithm, which continuously cycles through the updates and then performs a Trim-mean. And in particular, the algorithm uses Krum for selection. Thus Bulyan is a combination of Krum and Trim-mean. Specifically, Bulyan first iteratively uses Krum to select $\delta$ $(\delta\leq m-2c)$ local models. Then, Bulyan uses pruning averaging to aggregate $\delta$ local models. In particular, for each $j$-th model parameter, Bulyan ranks the $j$-th parameter of $\delta$ local models, finds $\gamma$ $(\gamma\leq\delta-2c)$ parameters that are closest to the median and calculates its mean as the $j$-th parameter of the global model. When $c\leq\frac{m-3}{4}$, Bulyan has theoretical guarantees for convergence of certain objective functions.

\subsubsection{Learned Lessons}
Existing aggregation methods, such as Trim-mean and Median, do not guarantee fidelity and robustness well\cite{cao2020fltrust}. In addition, Krum and Bulyan do not satisfy the efficiency goal because they require the server to compute the pairwise distances of local model updates for clients, which is computationally expensive when the number of clients is large. Bulyan is not scalable because it performs Krum multiple times in each iteration to calculate the pairwise distances between local models. Since the Euclidean distance between two local models may be influenced by individual model parameters, Krum may be affected by some anomalous model parameters\cite{guerraoui2018hidden}.


\section{Challenges and Future Directions}\label{challenge}
Although there are many ways to defend against model poisoning attacks, many problems still exist and need to be solved. Also, the existing methods are not effectively against all model poisoning attacks. For instance, the attack strategy against a robust FL proposed in \cite{cao2020fltrust} can pose a threat to most existing defense methods. We consider that a good defense mechanism should meet three requirements: 1) requires effective resistance to attacks; 2) resource conservation; and 3) ensuring data privacy.
In this section, we will discuss potential challenges and future research directions for defending against model poisoning attacks in FL.

\subsection{Resistance Effectiveness against Model Poisoning Attacks}

If an attacker makes obvious changes to updates, such as the appearance of extreme values, then such an attack is easily detected. However, there are few existing studies focusing on how to resist well-designed malicious updates. For example, an attacker can design an attack based on a generative adversarial network (GAN)\cite{zhang2019poisoning,hardy2019md} that makes it difficult for modified updates to be detected by the server.  In addition, an attacker can control multiple devices at the same time, or multiple malicious devices conspire to launch an attack. In this case, the contamination of local model updates can be adjusted according to the aggregation method of the global model.


In the future research, we need to be aware of well-designed model poisoning attacks. On the one hand, we will only study against general types of attacks, i.e., attacks that are stochastic and synchronous. On the other hand, we need to study the impact of different attack strategies, such as the number of malicious clients, the number of rounds and the time to launch the attack, on the effectiveness of the attack, so that we can design effective defense mechanisms.

\subsection{Computational Consumption of the Central Server}

The deployment of defense mechanisms in FL requires a certain amount of computational resources, which should not exceed the capacity of the central server. The existing defense mechanisms generally fail to explicitly consider the limitation of computational resources. For example, some verification mechanisms verify all the updates, which will not only consume energy but also result in time delay, thus affecting the whole FL training process. Moreover, we also need to consider the energy consumption of the server if it is required to collect data and train the model.

In future research, we need to consider how to reduce the resource consumption caused by deploying the defense mechanism. For FL with a small number of clients, the submitted local models can be verified one by one, but once the number of clients is huge, this can consume a lot of time and energy. One possible idea is to reduce resource consumption by designing FL with multiple servers, spreading the task of verifying updates to those servers. However, this design introduces new problems, such as communication cost and privacy leakage. The combination of blockchain and FL might be another promising solution \cite{wang2021blockchain,hu2021blockchain,chen2021robust}. In \cite{chen2021robust}, a blockchain-based FL is proposed to defend against malicious attacks. In this framework, clients upload updates to verifiers, who will select benign updates by voting, and then the selected updates will be aggregated and written to blocks through the blockchain network.



\section{Conclusion}\label{concl}
In this survey, we first investigate the existing defense methods against model poisoning attacks in FL, and then classify these methods into two main categories: evaluating local model updates and designing global aggregation model algorithms. We also analyze the challenges and future research directions regarding the model poisoning attacks in FL.



\bibliographystyle{unsrt}
\bibliography{reference.bib}

\end{document}